\documentclass[12pt,a4paper]{article}

\usepackage{amsmath}
\usepackage{amssymb}
\usepackage{hyperref}
\usepackage{graphicx}

\newcommand{\Lagr}{\mathcal{L}}
\newcommand{\tsum}{\mathop{\textstyle\sum}}
\newcommand{\mmass}[2][2]{{m_{#2}^{#1}}}
\newcommand{\masseps}[1][2]{\mmass{\epsilon}}
\newcommand{\masstop}[1][2]{\mmass{0}}
\newcommand{\cbeta}[2]{\beta_{#1}^{(#2)}}
\newcommand{\coeffv}[2]{v_{#1}^{(#2)}}
\newcommand{\cvtwid}[2]{\tilde{v}_{#1}^{(#2)}}
\newcommand{\eV}{\,\mathrm{eV}}
\newcommand{\keV}{\,\mathrm{keV}}
\newcommand{\MeV}{\,\mathrm{MeV}}

\newcommand{\deltaph}{\lambda}
\newcommand{\newop}[2]{\def#1{\mathop{\mathrm{#2}}\nolimits}}
\newop{\artanh}{artanh}

\newcommand{\sfrac}[2]{{\textstyle\frac{#1}{#2}}}
\newcommand{\indup}[1]{_{\scriptscriptstyle\mathrm{#1}}}

\newcommand{\decay}{F}

\newcommand{\pn}{\tilde{\pi}^0}

\newcommand{\equref}[1]{Eq.~(\ref{#1})}

\begin{document}

\hfill 

\hfill 

\bigskip\bigskip

\begin{center}

{{\Large\bf  Hadronic decays of $\mbox{\boldmath$\eta$}$ and
$\mbox{\boldmath$\eta'$}$ \\
with coupled channels
\footnote{Work supported in part by the DFG}\par}}

\end{center}

\vspace{.4in}

\begin{center}
{\large N. Beisert\footnote{email: nbeisert@physik.tu-muenchen.de}
 and  B. Borasoy\footnote{email: borasoy@physik.tu-muenchen.de}}

\bigskip\bigskip

\href{http://www.ph.tum.de/}{Physik-Department},
Technische Universit{\"a}t M{\"u}nchen\\
D-85747 Garching, Germany
\end{center}

\vspace{.7in}

\thispagestyle{empty} 

\begin{abstract}
The hadronic decays $\eta \rightarrow \pi \pi \pi$, $\eta' \rightarrow \pi \pi \pi$ and
$\eta' \rightarrow \eta \pi \pi$ are investigated within a $U(3)$ chiral unitary
approach. Final state interactions are included
by deriving the effective $s$-wave potentials for meson meson scattering
from the chiral effective Lagrangian and iterating them in a Bethe-Salpeter
equation.
With only a small set of parameters we are able to explain both rates
and spectral shapes of these decays.
\end{abstract}\bigskip

\begin{center}
\begin{tabular}{ll}
\textbf{PACS:}&12.39.Fe, 13.25.Jx\\[6pt]
\textbf{Keywords:}& Chiral effective lagrangians, hadronic decays of mesons, \\&
$\eta$ and $\eta'$, coupled channels, final state interactions.
\end{tabular}
\end{center}

\vfill

\section{Introduction}\label{sec:intro}
The hadronic decays $\eta \rightarrow \pi \pi \pi$ and 
$\eta' \rightarrow \pi \pi \pi$ are of great interest since they
violate isospin symmetry. At such low energies the calculations of the decay amplitudes
are based on
effective chiral Lagrangians which have the same symmetries and symmetry breaking 
patterns as the underlying QCD Lagrangian. They are  written in terms of
effective degrees of freedom, usually the octet of pseudoscalar mesons -- pions,
kaons and the eta~--~\cite{Gasser:1985gg}, however, they can easily be extended to
include the $\eta'$ also \cite{Kaiser:2000gs}. With the effective Lagrangian at hand,
one can perform a perturbative expansion of the decay amplitudes in powers
of the pseudoscalar meson masses and external momenta.
For the decay $\eta \rightarrow \pi \pi \pi$ the expansion parameters are the
ratios of the Goldstone boson octet masses or Lorentz invariant combinations of 
external momenta over the scale of spontaneous chiral symmetry breaking, 
$m_P^2/\Lambda_\chi^2 $ and  $(p_P \cdot p_{P'})/\Lambda_\chi^2 $, respectively,
with $\Lambda_\chi = 4 \pi f_\pi \approx 1.2$ GeV and
higher loops contribute to higher chiral orders.
At low energies these ratios are considerably smaller than unity and the chiral
expansion is expected to converge.

The inclusion of the $\eta'$, on the other hand,
spoils the conventional chiral counting scheme, since its mass $m_{\eta'}$ 
does not vanish in the chiral limit so that higher loops will still contribute to
lower chiral orders.
This can be prevented by imposing large $N_c$ counting rules
within the effective theory in addition to the chiral counting scheme.
In the large $N_c$ limit the axial anomaly vanishes and the $\eta'$ converts 
into a Goldstone boson. The properties of the theory may then be analyzed in a 
triple expansion in powers of small momenta, quark masses and 1/$N_c$, see
e.g. \cite{Kaiser:2000gs,Leutwyler:1996sa, Herrera-Siklody:1997pm, Kaiser:1998ds}. 
In particular, $m_{\eta'}$ is treated as a small 
quantity. 
Phenomenologically, this is not the case since $m_{\eta'} =958$ MeV, and we will therefore treat
the $\eta'$ as a massive state leading to a situation analogous to
chiral perturbation theory (ChPT) with baryons. Recently, a new regularization scheme -- the so-called
infrared regularization -- has been proposed which maintains Lorentz and chiral
invariance explicitly at all stages of the calculation while providing
a systematic counting scheme for the evaluation of the chiral loops \cite{Becher:1999he}.
Infrared regularization can be applied in the presence of any massive state
and has been employed in chiral perturbation theory including the $\eta'$ ($U(3)$ ChPT) 
in \cite{Borasoy:2001ik, Beisert:2001qb, Beisert:2002ad}, but its applicability
is restricted to very small external three momenta. This is surely not the case
for the decay $\eta' \rightarrow \pi \pi \pi$ where large unitarity corrections
are expected due to final state interactions between the three pions.

Final state interactions have already been shown to be substantial in
$\eta \rightarrow \pi \pi \pi$ both in a complete one-loop calculation
in $SU(3)$ ChPT \cite{Gasser:1985pr} and under the inclusion of final state interactions
using extended Khuri-Treiman equations \cite{Kambor:1996yc}. The last calculation
comes closest to the experimental decay rate but still remains below it,
if the usual quark mass ratios are employed. However, the experimental
rate of the decay can be reproduced by increasing the quark mass ratio
$[(m_d-m_u)/(m_s-\hat{m})] \cdot [(m_d+m_u)/(m_s+\hat{m})] $ from
$1/(24.1)^2$ to $1/(22.4 \pm 0.9)^2$ which is now considered to be the
most accurate determination of this quark mass ratio.
%
%
It is therefore obligatory to include final state interactions also
in the decay $\eta' \rightarrow \pi \pi \pi$.
We will  apply the following approach in the present investigation:
after deriving the effective potentials for meson meson scattering
from the chiral effective Lagrangian, we iterate them utilizing a Bethe-Salpeter
equation (BSE). This method has been proven to be useful both in
the purely mesonic sector and under the inclusion of baryons \cite{Oller:1997ti,Kaiser:1995eg}. 
The BSE generates dynamically bound states of the mesons and baryons and 
accounts for the exchange of resonances without including them explicitly.
The usefulness of this approach lies in the fact that from a small set of
parameters a large variety of data can be explained.
It has been extended recently to $U(3)$ ChPT in \cite{Beisert:2001A2} where
effects of the $\eta'$ in  meson meson scattering are investigated and the parameters
of the $U(3)$ chiral effective Lagrangian are constrained by comparing the results
with the experimental phase shifts. Again, with only a few chiral parameters
the phase shifts have been reproduced. Using the same technique, the electroproduction
of the $\eta'$ meson on nucleons has also been investigated yielding good agreement
with data \cite{Borasoy:2002EM}.
This method allows us to account for final state interactions and to study
the importance of resonances in the different isospin channels, a topic
of interest, e.g., in the dominant hadronic decay mode of the $\eta'$,
$\eta' \rightarrow \eta \pi \pi$. A full one-loop calculation of this
decay has been performed in \cite{Beisert:2002ad} utilizing infrared regularization
and reasonable agreement with experimental data was achieved.
However, both higher order final state interactions and the explicit
inclusion of resonances have been omitted.
It has been claimed, on the other hand, that this decay can be described in a 
tree-level model via the exchange of the
scalar mesons $\sigma(560), f_0(980)$ and $a_0(980)$ which are combined
together with $\kappa(900)$ into a nonet \cite{Fariborz:1999gr} 
(see also \cite{Schechter:1971tc,Singh:1975aq,Deshpande:1978iv,Bramon:1980ni,Castoldi:1988dm}).
The authors find that the exchange of the scalar resonance $a_0(980)$ 
dominates. By iterating the effective chiral potentials to infinte order
in a Bethe-Salpeter equation, we can investigate the importance of these resonances
explicitly.
Hence, the present work provides a unified approach to the  decays 
$\eta \rightarrow \pi \pi \pi$, $\eta' \rightarrow \pi \pi \pi$ and
$\eta' \rightarrow \eta \pi \pi$ with chiral symmetry and unitarity being 
the main ingredients. With only a few chiral parameters we can compare
our results with a variety of experimental data and even predict the decay rate
for $\eta' \rightarrow \pi^0 \pi^+ \pi^-$.
For simplicity, we will restrict ourselves to $s$-waves since good
agreement with experimental data is already obtained; an extension to higher
multipoles is straightforward.

In the next  section, we present the chiral effective Lagrangian up to fourth chiral order
and the $\pi^0 \eta \eta'$
mixing for different up- and down-quark masses $m_u \ne m_d$ is discussed in detail
as it arises from the second and fourth order Lagrangian, respectively.
In Section~\ref{sec:BSE}
the implementation of the final state interactions via the BSE is illustrated.
In Sections~\ref{sec:eta3pi} to \ref{sec:eta3pitree} the decays
$\eta \rightarrow \pi \pi \pi$, $\eta' \rightarrow \pi \pi \pi$ and
$\eta' \rightarrow \eta \pi \pi$ are discussed.
Our results are compared with experimental data.

\section{The Lagrangian density}\label{sec:lagr}

In this section the Lagrangians at second and fourth chiral order
are presented and the resulting $\pi^0$-$\eta_8$-$\eta_0$ mixing
is discussed.
%
%
%
%
%
The effective Lagrangian for the pseudoscalar meson nonet
($\pi, K, \eta_8, \eta_0$) reads up to second order in the derivative expansion
\cite{Kaiser:2000gs,Kaiser:1998ds,Borasoy:2001ik}
\begin{equation}  \label{eq:mes1}
{\cal L}^{(0+2)} = - V_0 
+V_1 \langle \partial_{\mu} U^{\dagger} \partial^{\mu}U \rangle  
+V_2 \langle U^\dagger \chi+\chi^\dagger U\rangle
+i V_3 \langle U^\dagger \chi-\chi^\dagger U\rangle
+V_4 \langle U^\dagger\partial^{\mu} U \rangle \langle U^\dagger\partial_{\mu} U \rangle,
\end{equation}
where $U$ is a unitary $3 \times 3$ matrix containing the Goldstone boson
octet $(\tilde\pi^\pm,\pn,\tilde K^\pm,\tilde K^0, \eta_8)$ and the $\eta_0$. 
Its dependence on $\pn, \eta_8$ and $\eta_0$ is given by
\begin{equation}
U=\exp\bigl(\mathrm{diag}(1,-1,0)\cdot i\pn/ \decay
+\mathrm{diag}(1,1,-2)\cdot i\eta_8/ \sqrt{3}\decay
+i\sqrt{2}\eta_0/\sqrt{3}\decay+\ldots\bigr).
\end{equation}
The expression $\langle \ldots \rangle$ denotes 
the trace in flavor space, $\decay$ is the pion decay constant in the chiral limit
and the quark mass matrix $\mathcal{M} = \mbox{diag}(m_u,m_d,m_s)$
enters in the combination  $\chi  = 2 B \mathcal{M} $
with $B = - \langle  0 | \bar{q} q | 0\rangle/ \decay^2$ being the order
parameter of the spontaneous symmetry violation.
As we do not consider external (axial-) vector currents in the present investigation,
the covariant derivatives have been replaced by partial ones.

The coefficients $V_i$ are functions of $\eta_0$,
$V_i(\eta_0/\decay)$,
and can be expanded in terms of this variable. At a given order of
derivatives of the meson fields $U$ and insertions of the quark mass matrix 
$\mathcal{M}$ one obtains an infinite string of increasing powers of 
$\eta_0$ with couplings which are not fixed by chiral 
symmetry.\footnote{Note that we do not make use of 1/$N_c$ counting rules.}
Parity conservation implies that the $V_i$ are all even functions
of $\eta_0$ except $V_3$, which is odd, and
$V_1(0) = V_2(0) = V_1(0)-3V_4(0)=\frac{1}{4}\decay^2$
 gives the correct  normalizaton
for the quadratic terms of the mesons.
The potentials $V_i$ are expanded in the singlet field $\eta_0$ 
\begin{eqnarray}\label{eq:vexpand}
V_i\Big[\frac{\eta_0}{\decay}\Big] &=& \coeffv{i}{0} + \coeffv{i}{2} 
\frac{\eta_0^2}{\decay^2} +
\coeffv{i}{4} \frac{\eta_0^4}{\decay^4} + \ldots
\qquad \mbox{for} \quad i= 0,1,2,4 \nonumber \\
V_3\Big[\frac{\eta_0}{\decay}\Big] &=& \coeffv{3}{1} \frac{\eta_0}{\decay} + 
\coeffv{3}{3} \frac{\eta_0^3}{\decay^3}
+ \ldots \quad 
\end{eqnarray}
with expansion coefficients $\coeffv{i}{j}$ to be determined phenomenologically.

In order to describe the isospin violating decays $\eta \rightarrow \pi \pi \pi$ and 
$\eta' \rightarrow \pi \pi \pi$, we need to distinguish between the up- and 
down-quark masses, $m_u$ and $m_d$, which leads to $\pn$-$\eta_8$-$\eta_0$ mixing.
Taking the Lagrangian from \equref{eq:mes1}, diagonalization of the mass matrix
to first order in isospin breaking $m_u -m_d$ yields the mass eigenstates
$\pi^0, \eta, \eta'$ with
\begin{eqnarray}\label{eq:mix2}
\pn &=& \pi^0 - \epsilon \eta - 2\epsilon\vartheta \eta'\nonumber \\ 
\eta_8 &=&  \epsilon \pi^0 + \eta - \vartheta \eta' \nonumber \\ 
\eta_0 &=&  3\epsilon \vartheta  \pi^0 + \vartheta \eta + \eta'  .
\end{eqnarray}
The angles $\epsilon$ and $\vartheta$
arise from the mass differences of the light quark masses $m_u, m_d, m_s$.
While $\vartheta$ breaks $SU(3)$ symmetry due to different strange and nonstrange
quark masses, $\epsilon$ is proportional to strong
isospin violation $m_d-m_u$
\begin{eqnarray}\label{eq:mixpar}
\epsilon &=&  \frac{\sqrt{3}}{4} \frac{m_d - m_u}{m_s-\hat m} 
=\frac{m^2_{\epsilon}}{\sqrt{3}\,(m^2_\eta-m^2_\pi)}
\nonumber \\ 
\vartheta &=& \frac{4\sqrt{2}\,\cvtwid{2}{1}}{3v_0^{(2)}} B (m_s-\hat m)
=\frac{\sqrt{2}\,\cvtwid{2}{1}}{v_0^{(2)}} (m^2_\eta-m^2_\pi)
\end{eqnarray}
where we used the abbreviations 
\begin{eqnarray} \label{eq:cvtwid21}
\cvtwid{2}{1} &=& \sfrac{1}{4}\decay^2-\sfrac{1}{2}\sqrt{6}\coeffv{3}{1} \nonumber\\
m^2_{\epsilon}&=& B (m_d - m_u)\nonumber\\
\hat m&=& \sfrac{1}{2} (m_d+m_u).
\end{eqnarray}
The quantity $m^2_{\epsilon}$ can be expressed in terms of physical meson masses
by applying Dashen's theorem \cite{Dashen:1969eg}, which implies the identity of the pion
and kaon electromagnetic mass shifts up to ${\cal O}(e^2p^2)$
\begin{equation}
m^2_{\epsilon} = m^2_{K^0}-m^2_{K^\pm}+m^2_{\pi^\pm}-m^2_{\pi^0} .
\end{equation}
At leading order in the chiral expansion the meson masses are given by 
\begin{eqnarray}
m^2_\pi&=&2B\hat m,\nonumber\\
m^2_{K^\pm}&=&B(m_s+m_u),\nonumber\\
m^2_{K^0}&=&B(m_s+m_d) .  
\end{eqnarray}
Isospin violation is known to be small,
hence $\epsilon$ is a small quantity despite being of
zeroth order in the quark masses. 
Terms of order $(m_d-m_u)^2$ are tiny and therefore neglected throughout.
In the isospin limit of equal up- and down-quark masses $m_u = m_d$ the angle
$\epsilon$ vanishes and $\pn$ does not undergo mixing
with $\eta_8$ and $\eta_0$.

However, this is not the whole story. As shown in \cite{Beisert:2001qb} the fourth order
Lagrangian contributes to $\eta$-$\eta'$ mixing at leading order as well
which is due to the fact that we count the $\eta'$ mass as quantity of zeroth
chiral order.
The fourth order Lagrangian has the form
\begin{equation} \label{eq:mes2}
\Lagr^{(4)}=\tsum\nolimits_i \beta_i(\eta_0)\, O_i
\end{equation}
with the fourth order operators
\begin{equation}
\begin{array}{ll} \label{eq:mixop}
O_{4\phantom{0}}=-\langle C^\mu C_\mu\rangle\langle M\rangle, \qquad &
O_{5\phantom{0}}=-\langle C^\mu C_\mu M\rangle,\\
O_{17}=\langle C^\mu \rangle\langle C_\mu\rangle\langle M\rangle, \qquad &
O_{18}=-\langle C^\mu \rangle\langle C_\mu M\rangle,\\
O_{6\phantom{0}}=\langle M\rangle\langle M\rangle,&
O_{7\phantom{0}}=\langle N\rangle\langle N\rangle,\\
O_{8\phantom{0}}=\sfrac{1}{2}\langle MM+NN\rangle,&
O_{12}=\sfrac{1}{4}\langle MM-NN\rangle,\\
\end{array}
\end{equation}
We used the abbreviations $C_\mu=U^\dagger \partial_\mu U$, 
$M=U^\dagger\chi+\chi^\dagger U$, $N=U^\dagger\chi-\chi^\dagger U$ and
the coefficients $\beta_i$ can be expanded in $\eta_0$ in the same manner 
as the $v_{i}$ in \eqref{eq:vexpand}.
The mixing between $\pn, \eta_8$ and $\eta_0$ cannot be described in terms
of just two angles any longer. Due to the operators in \equref{eq:mixop}
two subsequent transformations of the original fields
$\pn, \eta_8$ and $\eta_0$ are necessary to bring both the kinetic and mass terms
of the Lagrangian into diagonal form. At leading order in isospin breaking the new 
transformed fields are then related to the original ones by
%
\begin{eqnarray}\label{eq:mix4}
\pn &=&  (1+R_{\pn\pi^0}) \pi^0 + R_{\pn\eta}  \eta + R_{\pn\eta'}  \eta' \nonumber \\ 
\eta_8 &=&  R_{8\pi^0} \pi^0 + (1+R_{8\eta})  \eta + R_{8\eta'}  \eta' \nonumber \\ 
\eta_0 &=&  R_{0\pi^0} \pi^0 + R_{0\eta}  \eta + (1+R_{0\eta'})  \eta'  
\end{eqnarray}
with the mixing parameters given by 
%
\begin{equation}
\begin{array}{rlcrl}
R_{8\pi^0}^{(0)}&=\displaystyle\frac{\masseps}{\sqrt{3}(\mmass{\eta}-\mmass{\pi})},&\qquad&
  R_{\pn\eta}^{(0)}&=-R_{8\pi^0}^{(0)},\\[5pt]
  R_{8\pi^0}^{(2)}&=R_{8\pi^0}^{(0)}(R_{\pn\pi^0}^{(2)}+\sfrac{2}{3}\Delta\indup{GMO}),&&
  R_{\pn\eta}^{(2)}&=-R_{8\pi^0}^{(0)}(R_{8\eta}^{(2)}+\sfrac{2}{3}\Delta\indup{GMO}),\\[5pt]
R_{0 \eta}^{(2)}&=\displaystyle\frac{4\cvtwid{2}{1}(\mmass{\eta}-\mmass{\pi})}{\sqrt{2}F^2 \masstop},&&
  R_{8 \eta'}^{(2)}&=\displaystyle-R_{0 \eta}^{(2)}+\frac{8\cbeta{5,18}{0}(\mmass{\eta}-\mmass{\pi})}{\sqrt{2}F^2},\\[15pt]
R_{0\pi^0}^{(2)}&=3R_{8\pi^0}^{(0)}R_{0 \eta}^{(2)},&&
  R_{\pn\eta'}^{(2)}&=2R_{8\pi^0}^{(0)}R_{8 \eta'}^{(2)},  \\[5pt]
R_{\pn\pi^0}^{(2)}&=\displaystyle-\frac{4 \cbeta{5}{0} \mmass{\pi}+6 \cbeta{4}{0}(\mmass{\eta}+\mmass{\pi})}{F^2},&&
R_{0 \eta'}^{(2)}&=\displaystyle-\frac{2\cbeta{4,5,17,18}{0}(\mmass{\eta}+\mmass{\pi})}{F^2},\\[5pt]
  R_{8\eta}^{(2)}&=\displaystyle-\frac{4 \cbeta{5}{0} \mmass{\eta}+6 \cbeta{4}{0}(\mmass{\eta}+\mmass{\pi})}{F^2},
\end{array}
\end{equation}
where the superscript denotes the chiral order and we have employed the
abbreviations $\cbeta{4,5,17,18}{0}= 3 \cbeta{4}{0} + \cbeta{5}{0} - 9 \cbeta{17}{0} + 3 \cbeta{18}{0} $ 
and $\cbeta{5,18}{0}= \cbeta{5}{0} + 3 \cbeta{18}{0}/2$.
The quantity $m^2_{0}=2v_0^{(2)}/F^2$ is the mass of the $\eta'$ meson in the chiral limit and
the deviation from the Gell-Mann--Okubo mass relation for the pseudoscalar mesons is given by
\begin{equation}
\Delta\indup{GMO}=\frac{4\mmass{K}-\mmass{\pi}-3\mmass{\eta}}{\mmass{\eta}-\mmass{\pi}}=
\frac{6(\mmass{\eta}-\mmass{\pi})}{F^2}
\bigg[\cbeta{5}{0}-6\cbeta{8}{0}-12\cbeta{7}{0}+\frac{4(\cvtwid{2}{1})^2}{F^2 \masstop}\bigg].
\end{equation}

Finally, the remaining terms of the fourth order Lagrangian that we need in this
calculation are given by
\begin{equation}
\begin{array}{ll}
O_{0\phantom{0}}=\langle C^\mu C^\nu C_\mu C_\nu\rangle,&
O_{1\phantom{0}}=\langle C^\mu C_\mu\rangle\langle C^\nu C_\nu\rangle,\\
O_{2\phantom{0}}=\langle C^\mu C^\nu\rangle\langle C_\mu C_\nu\rangle,&
O_{3\phantom{0}}=\langle C^\mu C_\mu C^\nu C_\nu\rangle,\\
O_{13}=-\langle C^\mu\rangle\langle C_\mu C^\nu C_\nu\rangle,&
O_{14}=-\langle C^\mu\rangle\langle C_\mu\rangle \langle C^\nu C_\nu\rangle,\\
O_{15}=-\langle C^\mu\rangle\langle C^\nu\rangle \langle C_\mu C_\nu\rangle,\qquad&
O_{16}=\langle C^\mu\rangle\langle C_\mu\rangle\langle C^\nu\rangle\langle C_\nu\rangle,\\
O_{21}=\langle C^\mu C_\mu i N\rangle,&
O_{22}=\langle C^\mu C_\mu\rangle\langle i N\rangle,\\
O_{23}=\langle C^\mu \rangle\langle C_\mu i N\rangle,&
O_{24}=\langle C^\mu \rangle\langle C_\mu\rangle\langle i N\rangle,\\
O_{25}=\langle iMN\rangle,&
O_{26}=\langle M\rangle\langle iN\rangle.
\end{array}
\end{equation}
Usually the $\beta_0$ term is not presented in conventional ChPT, since
there is a Cayley-Hamilton matrix identity that enables one to 
remove this term leading to modified coefficients 
$\beta_i$, $i=1,2,3,13,14,15,16$ \cite{Herrera-Siklody:1997pm},
but for our purposes
it turns out to be more convenient to include it.
Hence, we do not make use of the Cayley-Hamilton identity
and keep all couplings in order to present the most general expressions
in terms of these parameters. One can then drop one of the 
$\beta_i$ involved in the Cayley-Hamilton identity
at any stage of the calculation.

\subsection{Values of LECs}\label{sec:lec}

The unknown coupling constants of the chiral Lagrangian -- the so-called low-energy
constants (LECs) -- need to be fit 
to experimental data. This has already been accomplished in \cite{Beisert:2001A2},
where after applying the same approach as in the present investigation
we constrained the LECs of the Lagrangian up to fourth chiral order by comparing
the results with the experimental phase shifts of meson meson scattering.
Agreement was achieved in the isospin $I=0, \sfrac{1}{2}$ channels up to 1.2 GeV and in the isopin 
$I=\sfrac{3}{2}, 2$ channels up to 1.5 GeV. The discrepancey to the data above 1.2 GeV
in the $I=0$ channel, e.g., is due
to the omission of the  $4\pi$ channel, see \cite{Kaminski:1997gc}.
In order to obtain reasonable agreement with the experimental phase shifts,
the values of the involved chiral parameters are then constrained to the following
ranges 
\begin{equation}
\label{eq:par}
\begin{array}{lll}
\cbeta{0}{0}=(0.6\pm 0.1)\times 10^{-3},&\quad&
\cbeta{3}{0}=(-0.5\pm 0.1)\times 10^{-3},
\\
\cbeta{5}{0}=(1.4\pm 0.2)\times 10^{-3},&\quad&
\cbeta{8}{0}=(0.2\pm 0.2)\times 10^{-3}.
\end{array}
\end{equation}
while keeping $v_1^{(0)} = v_2^{(0)} = \sfrac{1}{4} \decay^2 $
and all remaining parameters set to zero. 
However, variations of some of the parameters which have been set to zero
do not yield any significant effect for the phase shifts, and in principle they can
have finite values. Such a parameter is $\beta_6^{(0)}$ or the combination 
$\cvtwid{2}{1}$ from \equref{eq:cvtwid21}, and we have chosen $\beta_6^{(0)}=0$ and
$\cvtwid{2}{1}= v_2^{(0)} = \sfrac{1}{4} \decay^2$ in
\cite{Beisert:2001A2} but we are free to change their values in the present investigation
as long as they do not contradict the data for the phase shifts. As a matter of fact,
this work provides a test whether the same values for the LECs can be 
employed when describing the hadronic decays of the $\eta$ and $\eta'$ and will
allow for finetuning of the parameters that were not constrained in
\cite{Beisert:2001A2}, e.g. $\beta_6^{(0)}$ and $\cvtwid{2}{1}$.
A good fit to the decays discussed in this work is given by
\begin{equation}\label{eq:par2}
\begin{array}{lll}
\cvtwid{2}{1}= \cvtwid{2}{2}= 0 ,&&
\\
\cbeta{0}{0}=0.56\times 10^{-3},&\quad&
\cbeta{3}{0}=-0.3\times 10^{-3},
\\
\cbeta{5}{0}=1.4\times 10^{-3},&\quad&
\cbeta{6}{0}=0.06\times 10^{-3}.
\end{array}
\end{equation}
%
%
with $\cvtwid{2}{2}= \sfrac{1}{4}\decay^2-\sqrt{6}\coeffv{3}{1}-3\coeffv{2}{2}$ 
and all the remaining parameters being zero. Note that there have been moderate
changes with respect to the values in \equref{eq:par} which were necessary to obtain better
agreement with the experimental decay rates and the spectral shapes of the
Dalitz distributions. It is surprising that with a small number
of parameters we are able to explain a variety of data within the approach.

It is also important to note, that the parameter choice in \equref{eq:par2} is not unique, since
variations in one of the parameters may be compensated by the other ones.
Nevertheless, we prefer to work with this choice; it is capable of describing
the experimental phase shifts and the hadronic decays of the $\eta$ and $\eta'$ 
-- as we will see shortly -- with a minimal set of parameters and motivated
by the assumption that most of the OZI-violating and 1/$N_c$-suppressed parameters are not important,
although $\beta_6^{(0)}$ and $\coeffv{3}{1}, \coeffv{2}{2}$ in $\cvtwid{2}{1}, \cvtwid{2}{2}$
have small but non-vanishing values.

\section{Final state interactions}\label{sec:BSE}

\begin{figure}
\centering\includegraphics{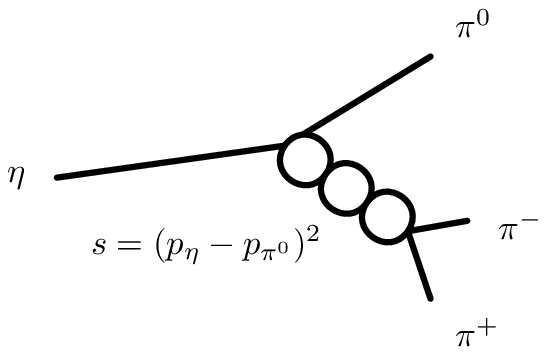}

\caption{Shown is a possible contribution to final state interactions in the decay 
$\eta\to\pi^0\pi^+\pi^-$. 
Here, it takes place in the $s$-channel between $\pi^+\pi^-$ and
the bubble chain represents part of the solution
to the BSE at energy $\sqrt{s}$.
}
\label{fig:Method}
\end{figure}

An investigation of the hadronic decay $\eta\to \pi\pi\pi$ 
at leading order in ChPT \cite{Cronin:1967jq, Osborn:1970nn} yields a decay width which
is significantly below the measured value \cite{Groom:2000in}. 
The result is improved by performing a next-to-leading order calculation,
but still fails to reproduce the phenomenological value.
The substantial increase from leading to next-to-leading order
already indicates that higher order effects beyond one-loop ChPT
may be important as well and should not be neglected. 
Kambor, Wiesendanger and Wyler \cite{Kambor:1996yc} were able 
to approximate these higher order effects by using Khuri-Treiman equations, which 
describe final state interactions. 
The underlying idea is that the initial particle, i.e. the $\eta$ or $\eta'$,
decays via chirally constrained vertices derived from the effective Lagrangian
into three mesons and that two out of these three mesons rescatter 
an arbitrary number of times,
see Fig. \ref{fig:Method} for illustration. There are three possible
ways in combining two of the mesons to a pair while leaving the third one unaffected
which corresponds to the $s$-, $t$- and $u$-channel, respectively.
Interactions of the third meson with the pair of rescattering mesons are neglected.
Such an infinite meson meson rescattering can alternatively be 
generated by application of the Bethe-Salpeter equation.
To this end, the $s$-wave potentials for meson meson scattering are derived from
the chiral effective Lagrangian and iterated in a Bethe-Salpeter equation.
The BSE then generates the propagator for two interacting particles in a covariant
fashion.

The Bethe-Salpeter equation for the two-particle propagator $T$ 
from a local interaction $A$ is given by 
\cite{Beisert:2001A2,Nieves:2000bx} 
\begin{equation}\label{eq:ResBSElong}
T(p,q_i,q_f)=A(p,q_i,q_f)+
\int\frac{i\,d^d k}{(2\pi)^d}\,\frac{T(p,q_i,k) A(p,k,q_f)}{(k^2-m^2)(\bar k^2-\bar m^2)}.
\end{equation}
Combinatorial factors for two identical particles in the loop have to be taken 
into account, but we prefer to keep this form of the BSE and 
modify the amplitudes accordingly.
In that case $A$ must be the four-point amplitude from 
ChPT multiplied by $-\frac{1}{2}$ in order for
$T$ to be proportional to a bubble chain with
the correct factors from perturbation theory. The factor of
$\frac{1}{2}$ is the symmetry factor of two identical
particle multiplets in a loop and $-1$ stems from factors of $i$ in the
vertices and propagators.

\begin{figure}
\[
\parbox{2cm}{\centering\includegraphics[scale=0.7]{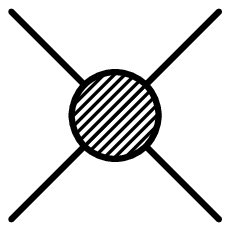}}
=
\parbox{2cm}{\centering\includegraphics[scale=0.7]{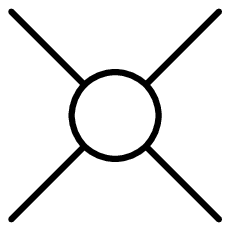}}
+
\parbox{3.25cm}{\centering\includegraphics[scale=0.7]{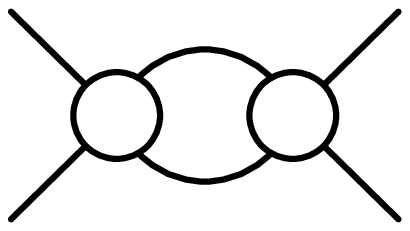}}
+
\parbox{4.5cm}{\centering\includegraphics[scale=0.7]{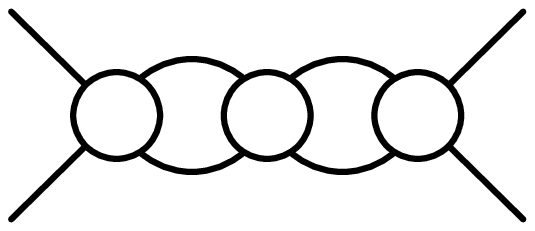}}
+\ldots
\]
\caption{Solution to the Bethe-Salpeter equation 
$T=A+AGA+AGAGA+\ldots$}
\label{fig:SolBSE}
\end{figure}

The solution for $T$ amounts to an infinite resummation of the interaction
kernels $A$ and the free propagators $G$ and can be diagrammaticaly depicted
as an infinite interaction chain, Fig. \ref{fig:SolBSE}.
In a short notation, both the equation and its solution can 
be written as 
\begin{equation}\label{eq:BSEMatrix}
T=A+TGA,\qquad T=(1-GA)^{-1}A=A+AGA+AGAGA+\ldots
\end{equation}
For the decays considered here we would like to use these
final-state interactions in such a way as to approximate to a large extent
the one-loop result from conventional ChPT.
The second term in the expansion of $T$, $AGA$,
equals the unitary corrections in
one-loop ChPT (i.e. without tadpoles). Tadpoles which are not included in our approach have been shown
to yield numerically small effects in scattering processes in the physical region
and can furthermore be partially absorbed by redefining chiral parameters \cite{Oller:1997ti}.
We have therefore neglected the tadpoles and crossed diagrams and restrict ourselves
to the tree diagrams from next-to-leading
order ChPT in the calculation of the potential $A$.
As mentioned above, for the decays one must take into account the three
possible ways in combining two of the light mesons to a pair while leaving the third one unaffected,
which corresponds to the $s$-, $t$- and $u$-channel.
Final state interactions will occur in all of these channels 
and in order to obtain the full amplitude one must add
an interaction chain in each channel,
$T_s+T_t+T_u$. 
This reproduces the unitary corrections at one-loop order;
the leading contribution in this sum, however, does not yield the tree level
result $A$ from ChPT. Portions of the tree level amplitude are included
in each $T$ and must be subtracted from the bubble chain.
The proper expression can be easily accounted for by choosing the  full
amplitude as $T=A+(T_s-A_s)+(T_t-A_t)+(T_u-A_u)$
which includes the unitary corrections in one-loop ChPT, while 
the leading contact term pieces $A_{s,t,u}$ are removed from the solutions 
$T_{s,t,u}$ and replaced by the tree diagram amplitude $A$.

In the evaluation of the BSE 
we will restrict ourselves to $s$-waves, since
they are expected to dominate in the decays discussed here, and indeed
this simplification gives results already in agreement with experiment. 
An improvement of the results -- particularly for the spectral shape 
of the Dalitz distributions --
may be expected by including the $p$-waves, but this is beyond the
scope of the present investigation.
We can further simplify the integral in the BSE \eqref{eq:ResBSElong}, since
we are only interested in the physically relevant piece of the solution $T$
with all momenta put on the mass shell.
The amplitude $A$ contains in general off-shell parts which deliver 
via the integral a contribution even to the on-shell part of the 
solution $T$. However, these off-shell parts yield exclusively chiral logarithms 
which -- besides being numerically small -- can be absorbed
by redefining the regularization scale of the loop integral. 
Furthermore the off-shell parts are not uniquely defined in 
ChPT.
We will therefore set all the momenta in the amplitudes in \eqref{eq:ResBSElong}
on-shell, which has also been done in previous work, see e.g. \cite{Oller:1997ti}. 
Finally, we need to consider all two-meson channels. In particular,  
those which have the same set of quantum numbers can couple amongst each other.
Consequently, the objects $A,T,G$ are promoted to matrices in channel space
and are functions of the two particle invariant mass squared $p^2$. 
With these simplifications the BSE equation 
reduces to \eqref{eq:BSEMatrix} as a matrix equation 
for every $p^2$ and the solution is obtained by matrix inversion.
The matrix $G$ is a diagonal matrix with elements 
given by the integral 
\begin{equation}
G_{m \bar{m}}(p^2) =\mathrel{}
\int\frac{\,d^d k}{(2\pi)^d}\,\frac{i}{(k^2-m^2)( (k-p)^2-\bar m^2)}
\end{equation}
where $m$ and $\bar m$ are the masses of the two mesons in the
corresponding channel.
In dimensional regularization the integral is evaluated as
{\arraycolsep0pt\begin{eqnarray}
\nonumber \\
G_{m \bar{m}}(p^2) =&&  \mathrel{}
\frac{1}{16\pi^2}\bigg[-1+
\ln\frac{m \bar{m}}{\mu^2}
  +\frac{m^2-\bar{m}^2}{p^2}\ln\frac{m}{\bar{m}}
\nonumber\\&&\qquad\qquad\quad\mathord{}
  -\frac{2\sqrt{\deltaph_{m\bar{m}}(p^2)}}{p^2}\artanh\frac{
  \sqrt{\deltaph_{m\bar{m}}(p^2)}}{(m+\bar{m})^2-p^2}\bigg]
\nonumber\\
\deltaph_{m\bar{m}}(p^2)=&&\mathrel{}\big((m-\bar{m})^2-p^2\big)\big((m+\bar{m})^2-p^2\big).
\end{eqnarray}}%
The integral is divergent and $\mu$ takes the role of a 
regularization constant which can be chosen independently
for every channel.
With these simplifications we write the full four-point amplitude $T_{abcd}$ as
a sum of the solutions $T_{ab,cd}$ of  the BSE in \equref{eq:ResBSElong}
in the $s$-, $t$- and $u$-channel
\begin{equation}\label{eq:4ptBSE}
T_{abcd}(s,t,u)=A_{abcd}(s,t,u)+(T-A)_{ab,cd}(s)+(T-A)_{ac,bd}(t)+(T-A)_{ad,bc}(u),
\end{equation}
where, e.g.,  $(T-A)_{ab,cd}(s)$ is a bubble chain 
in the $s$-channel with particles $ab$ on one side,
particles $cd$ on the other side, and the tree level piece removed.
The expression $(T-A)_{ab,cd}(s)$ is thus equivalent to the $s$-wave projection of the
final-state interactions of $T$, whereas the 
tree-level amplitude $A_{abcd}(s,t,u)$ is not decomposed into partial waves.

In the considered decays, $\eta,\eta'\to \pi\pi\pi$ and $\eta'\to\eta\pi\pi$,
the two-particle states are either uncharged or simply charged. 
There are nine uncharged combinations of mesons 
\begin{equation}
\pi^0\pi^0, \pi^+\pi^-,
\eta\pi^0, \eta\eta, 
K^0\bar K^0, K^+K^-,
\eta'\pi^0, \eta'\eta, \eta'\eta'
\end{equation}
which are labeled with indices $1, \ldots , 9$, respectively, 
and a set four charged channels
\begin{equation}
\pi^0\pi^+, \eta\pi^+, K^+\bar K^0, \eta'\pi^+ 
\end{equation}
labeled with indices $1, \ldots , 4$.
Due to isospin breaking, couplings between the different channels of a set occur,
but charge conservation prevents transitions between both sets. 
One obtains a $9\times 9$ matrix $A^0$ which summarizes the amplitudes
for the uncharged channels and a $4\times 4$ matrix $A^+$ for the 
charged channels. The division of the two-particle states into charged and
uncharged channels simplifies the treatment of the BSE.
We have presented the leading order contributions for $A^0$ and $A^+$ in the appendix.

In order to obtain the final expressions  $T_{abcd}$ for the decay amplitudes, the
solutions $T^0$ and $T^+$ of the BSE in \equref{eq:ResBSElong} must be 
corrected due to the overall symmetry factor $-1/2$ which was initially absorbed
into the matrix $A$. This is achieved by setting
\begin{equation}\label{eq:translate}
\begin{array}{rlcrl}
T_{\eta 0,00}&=-\sqrt{2}T^0_{31},&\qquad&
T_{\eta'0,00}&=-\sqrt{2}T^0_{71},\\
T_{\eta 0,+-}&=-T^0_{32},&&
T_{\eta'0,+-}&=-T^0_{72},\\
T_{\eta +,0+}&=-T^+_{21},&&
T_{\eta'+,0+}&=-T^+_{41},\\
T_{\eta'\eta ,00}&=-\sqrt{2}T^0_{81},&&
T_{\eta'0,\eta 0}&=-T^0_{83},\\
T_{\eta'\eta ,+-}&=-T^0_{82},&&
T_{\eta'+,\eta +}&=-T^+_{42},
\end{array}
\end{equation}
where an index $0,\pm$ denotes $\pi^{0}, \pi^\pm$. 
The same relations hold between the amplitudes $A_{ab,cd}$ and $A^{+,0}_{ij}$.
The additional factors $\sqrt{2}$ in some of the channels are included in order
to account for the statistical factor occuring in states with identical 
particles $(\pi^0, \pi^0)$.

\section{$\eta \rightarrow \pi \pi \pi$} \label{sec:eta3pi}

The decay $\eta \rightarrow \pi \pi \pi$ violates isospin and can 
take only place due to a finite mass difference $m_u-m_d$ or electromagnetic
interactions. The latter ones are expected to be small (Sutherland's theorem)
\cite{Sutherland:1966mi} which has been confirmed in an effective Lagrangian 
framework \cite{Kambor:1996yc}. 
Disregarding these, the amplitude is proportional to $m_u-m_d$ and provides
a sensitive probe on $SU(2)$ symmetry breaking.

The process $\eta \rightarrow \pi \pi \pi$ is the dominant decay mode of the $\eta$
with the measured rates being \cite{Groom:2000in}
\begin{eqnarray} \label{eq:decayrate}
\Gamma(\eta \rightarrow \pi^0 \pi^0 \pi^0)  &=& (379 \pm 40)\eV
\nonumber \\
\Gamma(\eta \rightarrow \pi^0 \pi^+ \pi^-)  &=& (274 \pm 33)\eV 
\end{eqnarray}
and the ratio given by
\begin{equation} 
r = \frac{\Gamma(\eta \rightarrow \pi^0 \pi^0 \pi^0)}
         {\Gamma(\eta \rightarrow \pi^0 \pi^+ \pi^-)} =  1.35 \pm 0.05
\end{equation}

Further information is provided by the energy dependence of the Dalitz plot
which must be compared with our results. 
To this end, we introduce a compact notation which can be applied to all decays
considered in the present work. Let $\phi_1 \rightarrow \phi_2 \phi_3 \phi_4$ be the decay of
a meson with mass $M$ into three lighter mesons with masses $m_2, m_3$ and $m_4$.
For the cases considered here, we will choose $m_3=m_4=m_{34}$.
The Dalitz plot distribution is conventionally described
in terms of the two variables
\begin{eqnarray} \label{eq:dalitz}
x &=&\frac{\sqrt{3}(u-t)}{2 M (M-m_2-2m_{34})} 
\nonumber \\
y &=&  \frac{(m_2+2m_{34})\big[ (M-m_2)^2 -s \big]}{2 M m_{34}(M-m_2-2m_{34})}  -1
\end{eqnarray}
with the Mandelstam variables
\begin{equation} 
s = (p_1 - p_2)^2, \qquad
t = (p_1 - p_3)^2, \qquad
u = (p_1 - p_4)^2 
\end{equation}
with $p_k$ and $m_k$ being the four-momentum and mass of the $k$-th meson and $p_1^2 =M^2$.
The Dalitz plot distribution $|A(x,y)|^2$ is then expanded as
\begin{equation} 
|A(x,y)|^2  = N\big[ 1 + a y + b y^2 + c x^2 \big].
\end{equation}
for the charged decay and
\begin{equation} 
|A(x,y)|^2  = N' \big[ 1 + g( y^2 + x^2) \big]
\end{equation}
for the neutral decay with no linear term in $y$ and only one coefficient $g$ due to
Bose symmetry.
The expansion parameters have been determined experimentally to be
\begin{equation} \label{eq:exp}
\begin{array}{llll}
\mbox{Layter et al. \cite{Layter:1973ti}:} \qquad &a= -1.08\pm 0.014,&b= 0.034\pm 0.027,\quad& c=0.046\pm 0.031\\
\mbox{Gormley et al. \cite{Gormley:1970qz}:}  \qquad &a= -1.17\pm 0.02,&b= 0.21\pm 0.03,& c=0.06\pm 0.04 \\
\mbox{Amsler et al. \cite{Amsler:1995sy}:}  \qquad &a= -0.94\pm 0.15,&b= 0.11\pm 0.27,& \\
\mbox{Amsler et al. \cite{Amsler:1998up}:}  \qquad &a= -1.19\pm 0.07,&b= 0.19\pm 0.11,& \\
\mbox{Abele et al. \cite{Abele:1998yj}:}  \qquad &a= -1.22\pm 0.07,&b= 0.22\pm 0.11,& c=0.06\mbox{ fixed}\\[0.1cm]
\mbox{Alde et al. \cite{Alde:1984wj}:}  \qquad &g= -0.044\pm 0.046,& & \\
\mbox{Abele et al. \cite{Abele:1998yi}:}  \qquad &g= -0.104\pm 0.04,& & \\
\mbox{Tippens et al. \cite{Tippens:2001fm}:}  \qquad &g= -0.062\pm 0.010 .\quad& & 
\end{array}
\end{equation}
%

\subsection{Tree diagram contributions to $\eta \rightarrow \pi \pi \pi$}
\label{sec:eta3pitree}

The first step in our approach is to calculate the tree diagram contributions
with vertices from the Lagrangians in Eqs.~(\ref{eq:mes1}, \ref{eq:mes2}).
The tree diagram contributions for $A_{\eta \rightarrow \pi^0 \pi^+ \pi^-}$
from ${\cal L}^{(0+2)}$ and ${\cal L}^{(4)}$ read
\begin{eqnarray}
A_{\eta0+-}(s,t,u)&= &
\frac{m^2_{\epsilon}(4m^2_\pi-3s)}{3\sqrt{3} F_\pi^2 (m^2_\eta-m^2_\pi)} 
\bigg[
1
+\frac{8m^2_\pi(F_K-F_\pi)}{3(m^2_\eta-m^2_\pi)F_\pi}
+\frac{2\Delta\indup{GMO}(8m^2_\pi-3s)}{3(4m^2_\pi-3s)}
\nonumber\\&&
\qquad\qquad+\frac{(4\beta_{0}^{(0)}-2\beta_{3}^{(0)})}{F_\pi^2(4m^2_\pi-3s)}
\big((t+u-s)^2-s^2-(t-u)^2\big)
\bigg],
\end{eqnarray}
where we have employed the expressions for the
meson masses $m_\phi$ and the decay constants
$F_\phi$ at fourth chiral order \cite{Beisert:2001qb}.
It is interesting to see that this expression from
$U(3)$ ChPT equals the one obtained in \cite{Gasser:1985pr} in $SU(3)$ ChPT. 
The differences due to $\eta\eta'$ mixing are completely absorbed
by a change in $\Delta\indup{GMO}$.
The expression for the neutral decay follows immediately
via the $\Delta I=1$ selection rule
\begin{equation} \label{eq:selrul}
A_{\eta 000}(s,t,u) =
A_{\eta 0+-}(s,t,u) +A_{\eta 0+-}(t,u,s) 
+A_{\eta 0+-}(u,s,t) .
\end{equation}
Taking the central values for the LECs from \equref{eq:par2} and applying
Dashen's theorem 
\begin{equation} \label{eq:dash}
B (m_d -m_u) =m^2_\epsilon=  m^2_{K^0} - m^2_{K^+} + m^2_{\pi^+} -m^2_{\pi^0} ,
\end{equation}
we obtain for the rates of the decays
\begin{equation} 
\Gamma(\eta \rightarrow \pi^0 \pi^+ \pi^-)  = 88 \eV , \qquad
\Gamma(\eta \rightarrow \pi^0 \pi^0 \pi^0 )  = 134 \eV 
\end{equation}
with the ratio given by $r= 1.52$. 
Clearly, the results are far
from being even close to the experimental value; this is well-known for quite some time
\cite{Cronin:1967jq, Osborn:1970nn}. A full one-loop calculation within
conventional ChPT \cite{Gasser:1985pr}, although improving the result considerably
($\Gamma = 167 \pm 50 \eV$),
still fails in being in agreement with the current data.
The unitarity corrections, particularly in the $\pi\pi$ channel, turn out
to be large and indicate that a summation of two-pion rescatterings is
necessary which cannot be done in a loopwise expansion.
Higher order corrections to Dashen's theorem as discussed in the literature
\cite{Donoghue:1992ha, Donoghue:1993hj, Bijnens:1993ae}
may increase the result a bit further, but cannot account for the large
discrepancy to the data. We will return to this point in the next section.

The resulting parameters of the 
Dalitz plot distribution are
\begin{equation} 
a= -1.14 , \quad  b= 0.31 , \quad c=0.014; \qquad g= 0.
\end{equation}
While the parameters for the charged decay seem to be in agreement, the coefficient
$g$ of the uncharged decay vanishes identically in this approximation and is in disagreement
with the recent precise measurement of \cite{Tippens:2001fm}.

\subsection{Final state interactions in $\eta \rightarrow \pi \pi \pi$}
\label{sec:eta3pifsi}

As already mentioned in the introduction, final state interactions
due to two-pion rescatterings in the decay $\eta \rightarrow \pi \pi \pi$
have been evaluated in \cite{Kambor:1996yc} using Khuri-Treiman equations and making
use of the one-loop ChPT result, see also \cite{Anisovich:1996tx,Bijnens:2002qy}
for a similar approach.
Our framework is much simpler than that and provides a unified aproach
to the hadronic decays of the $\eta$ and $\eta'$.
The underlying idea is that the initial particle, in the present case the $\eta$,
decays via chirally constrained vertices derived from the effective Lagrangian
into three mesons and that two out of these three mesons rescatter. There are three possible
ways in combining two of the mesons to a pair while leaving the third one unaffected
which corresponds to the $s$-, $t$- and $u$-channel, respectively.

Following 
\eqref{eq:4ptBSE} and \eqref{eq:translate}
in our approach, the full amplitudes read
\begin{eqnarray}
T_{\eta0+-}(s,t,u)&=&
A_{\eta0+-}(s,t,u)
-(T-A)_{32}^{0}(s)
-(T-A)_{21}^{+}(t)
-(T-A)_{21}^{+}(u)\\\nonumber
T_{\eta000}(s,t,u)&=&
A_{\eta000}(s,t,u)
-\sqrt{2}(T-A)_{31}^{0}(s)
-\sqrt{2}(T-A)_{31}^{0}(t)
-\sqrt{2}(T-A)_{31}^{0}(u).
\end{eqnarray}
Note that the $\Delta I=1$ selection rule, \equref{eq:selrul}, 
is not valid any longer, 
since our method includes higher orders in the isospin breaking parameter $m_u -m_d$
which spoil this relation.
The difference to this rule is therefore proportional to higher orders in $m_u -m_d$,
and we verified numerically that the deviations are rather small. 
Using the parameters from \equref{eq:par2} we obtain the results
\begin{equation} \label{eq:eta3pires}
\Gamma(\eta \rightarrow \pi^0 \pi^+ \pi^-)  = 237 \eV ,\qquad
\Gamma(\eta \rightarrow \pi^0 \pi^0 \pi^0)  = 351 \eV 
\end{equation}
with the ratio given by $r= 1.48$. Our decay rates are close to the 
experimental ones, and for the neutral decay we are even within the error bars.

We have employed Dashen's theorem \equref{eq:dash}, in order to arrive
at the results \equref{eq:eta3pires}, however, slightly modified values
for the quark mass ratios increase the result even further \cite{Kambor:1996yc}.
To this end, one replaces the quark mass difference $m_u - m_d$ by
\begin{equation} \label{eq:mu-md}
B (m_d -m_u) = \frac{1}{Q^2}\frac{m^2_{K}}{m^2_{\pi}} ( m^2_{K}   -m^2_{\pi})
\end{equation}
with
\begin{equation} 
Q^2 = \frac{m_s - \hat{m}}{m_d - m_u} \frac{m_s + \hat{m}}{m_d + m_u}
\end{equation}
Dashen's theorem yields the value
\begin{equation} 
Q^2_{\mathrm{Dashen}} = \frac{m^2_{K}}{m^2_{\pi}} 
 \frac{ m^2_{K}   -m^2_{\pi}}{m^2_{K^0} - m^2_{K^+} + m^2_{\pi^+} -m^2_{\pi^0}} = (24.1)^2 .
\end{equation}
Alternatively, the decay $\eta \rightarrow 3 \pi$ can be used to determine $Q$.
In order to reproduce the  experimental values of \equref{eq:decayrate}
we obtain
\begin{equation} 
Q = 23.4 \pm 0.8 , 
\end{equation}
a value which is in agreement with the requirement by Dashen's theorem and
indicates that higher order corrections to this low-energy theorem may be small,
in contradistinction to what has been discussed in 
\cite{Donoghue:1992ha, Donoghue:1993hj, Bijnens:1993ae}.

Let us turn now to the comparison of the spectral shapes for these decays.
Our method yields
\begin{equation} 
a= -1.24 , \quad  b= 0.37 , \quad c=0.05; \qquad g= -0.014 .
\end{equation}
The results are in the same ballpark as those from \cite{Kambor:1996yc} and
indicate that our method captures the same physics as the elaborate work of
\cite{Kambor:1996yc}. The situation is more complex when our
results are compared with the data in \equref{eq:exp}. The parameters
$a$ and $b$ are in agreement with \cite{Amsler:1998up},\cite{Abele:1998yj} $(a)$ 
and \cite{Amsler:1995sy} $(b)$, respectively, but
not with the other ones. Also, our value for $g$ is excluded by the most recent
measurement \cite{Tippens:2001fm}, but since the experimental uncertainties
are significant in most cases, a new determination of the spectral shape with higher statistics
is needed.

\subsection{Resonances in $\eta \rightarrow \pi \pi \pi$}\label{sec:eta3pires}

It is interesting 
to investigate which resonances could be of importance for the decay 
and yield a substantial contribution to the decay width. 
We are able to study the effects of the resonances within
our framework, since we can in principle switch on and off the final state interactions 
\eqref{eq:4ptBSE}
in the $s,t,u$ channels separatly.
For the neutral decay, however, this would destroy
Bose symmetry and we will restrict ourselves to the charged decays, for which we will switch
on the $t$- and $u$-channel only simultaneously, in order to preserve Bose symmetry.
The results for the decay $\eta\to\pi^0\pi^+\pi^-$ are
\begin{equation}\label{eq:eta3piResonances}
\begin{array}{lllll}
\mbox{tree:}&
\Gamma= 88 \eV , &
a= -1.14 ,&  b= 0.31 ,& c=0.014
\\
\mbox{tree}+s\mbox{-channel:}&
\Gamma  = 316 \eV ,\quad & a= -1.04 ,\quad& b= 0.23 , \quad& c=0.007\\
\mbox{tree}+t,u\mbox{-channel:}&\Gamma  = 50 \eV , &
a= -1.53 , &  b= 0.61 , & c=0.093
\\
\mbox{tree}+s,t,u\mbox{-channel:}\qquad&\Gamma  = 237 \eV , &
a= -1.24 , &  b= 0.37 , & c=0.005
\end{array}
\end{equation}
Two of the three decay products can form either an isoscalar or an isovector state
and their invariant mass must lie between 
$2 m_\pi \approx 275\MeV$ and $m_\eta-m_\pi \approx 410\MeV$. 
Furthermore, the $t,u$-channels are charged, hence
the $I=0$ resonances cannot propagate in these channels.
The so-called $\sigma$ resonance is in the appropriate mass range, 
while the resonances $f_0(980)$ and $a_0(980)$
may also influence the decay, although their masses
are comparatively large.
From \eqref{eq:eta3piResonances} one observes that 
final state interactions in the $t,u$-channel 
decrease the decay width by a small amount. 
In the resonance picture this could be interpreted as
the tail of the $I=1$ resonance $a_0(980)$. 
Large effects on the decay width stem from 
$s$-channel final state interactions
where also $I=0$ resonances are allowed. 
The $f_0(980)$ is as far away from 
the physical region as the $a_0(980)$ and is
thus expected to have a small impact on the decay, so that the 
enhancement in the decay width can be mainly attributed
to the $\sigma$.

\section{$\eta' \rightarrow \pi \pi \pi$}\label{sec:etap3pi}

We will now apply the same method to the decay $\eta' \rightarrow \pi \pi \pi$
in which the final state interactions should also play an important role.
According to our knowledge, there exist only tree level calculations for this decay
in which unitarity corrections have been neglected throughout 
\cite{DiVecchia:1981sq,Akhoury:1989ed,Fajfer:1989ij,Herrera-Siklody:1999ss}. All of them
disagree with the present experimental knowledge.
E.g., in the framework of large $N_c$ ChPT this decay was calculated up to 
next-to-leading order \cite{Herrera-Siklody:1999ss}
(loops start at next-to-next-to-leading order) yielding
the decay rates $\Gamma(\eta' \rightarrow \pi^0 \pi^0 \pi^0)  \approx 2200 \eV$
and $\Gamma(\eta' \rightarrow \pi^0 \pi^+ \pi^-) \approx 1600 \eV $
which are almost an order of magnitude larger than the data.
The measured rates of this decay mode are given by \cite{Groom:2000in}
\begin{equation} \label{eq:decayrate2}
\Gamma(\eta' \rightarrow \pi^0 \pi^0 \pi^0)  = (311 \pm 77)\eV,\qquad
\Gamma(\eta' \rightarrow \pi^0 \pi^+ \pi^-)  <1005 \eV 
\end{equation}
with the ratio 
\begin{equation} 
r = \frac{\Gamma(\eta' \rightarrow \pi^0 \pi^0 \pi^0)}
         {\Gamma(\eta' \rightarrow \pi^0 \pi^+ \pi^-)} >  0.3 .
\end{equation}
Hence, the theoretical situation is unsatisfactory requiring the inclusion of unitary corrections.

\subsection{Tree diagram contributions to $\eta' \rightarrow \pi \pi \pi$}
\label{sec:etap3pitree}

The tree diagram contributions
with vertices from the Lagrangians ${\cal L}^{(0+2)}$
in Eqs.~(\ref{eq:mes1}, \ref{eq:mes2}) read
for $A_{\eta' 0 + -}$
\begin{equation}
A_{\eta' 0 + -}(s,t,u) = 
-\frac{4\sqrt{6}\,m^2_\epsilon(m^2_\eta-2m^2_\pi)\cvtwid{2}{1}}{9(m^2_\eta-m^2_\pi)F^4}.
\end{equation}
The pertinent tree level amplitude for the neutral decay follows from the $\Delta I=1$ selection rule.
Employing Dashen's theorem, \equref{eq:dash}, leads to
\begin{equation} \label{eq:decetap3pi}
\Gamma(\eta' \rightarrow \pi^0 \pi^+ \pi^-)  = 114 \eV ,\qquad
\Gamma(\eta' \rightarrow \pi^0 \pi^0 \pi^0 )  = 55 \eV  
\end{equation}
with the ratio given by $r= 0.48$.
As expected, we cannot reproduce the experimental rate for the charged
decay, but it is worthwile mentioning that our result for the charged decay
remains below the experimental
decay rate in contradistinction to the rates in 
\cite{DiVecchia:1981sq,Akhoury:1989ed,Fajfer:1989ij,Herrera-Siklody:1999ss} 
ranging from 450 eV in \cite{Fajfer:1989ij} to 1600 eV in \cite{Herrera-Siklody:1999ss}.

The Dalitz plot parameters read in the tree diagram approximation
\begin{equation} 
a= -7.72 , \quad  b= 5.78 , \quad c=0.87; \qquad g= 0.94 .
\end{equation}
%
Unfortunaly, experimental data on the spectral shape is not available.

\subsection{Final state interactions in $\eta' \rightarrow \pi \pi \pi$}
\label{sec:etap3pifsi}

The smallness of our results in Eqs.~\eqref{eq:decetap3pi}
hints at the importance of unitary corrections. Applying the same model as 
in the previous section we arrive at
\begin{equation} \label{eq:etap3pires}
\Gamma(\eta' \rightarrow \pi^0 \pi^+ \pi^-)  = 294 \eV , \qquad
\Gamma(\eta' \rightarrow \pi^0 \pi^0 \pi^0)  = 291 \eV 
\end{equation}
which is in good agreement with the available data. In fact, the result
for the charged decay is a prediction and could serve as a check of our approach
in a future experiment. The ratio of the two decays is given by $r= 0.99$.
The slope parameters of the Dalitz plot are given by
\begin{equation} 
a= -3.08 , \quad  b= 0.13 , \quad c=0.62; \qquad g= -0.86 ,
\end{equation}
and need to be verified experimentally.
However, these predictions must be treated with care, since for these decays 
we expect contributions from $p$-waves as well due to the presence of the $\rho(770)$.

\section{$\eta' \rightarrow \eta \pi \pi$}\label{sec:etapeta2pi}

The last decay we are going to discuss in this work is the dominant decay
mode of the $\eta'$, $\eta' \rightarrow \eta \pi \pi$.
The measured rates are \cite{Groom:2000in}:
\begin{equation} 
\Gamma(\eta' \rightarrow \eta \pi^+ \pi^-)  = (90 \pm 8)\keV ,\qquad
\Gamma(\eta' \rightarrow \eta \pi^0 \pi^0)  = (42 \pm 4)\keV
\end{equation}
with the ratio given by
\begin{equation} 
r = \frac{\Gamma(\eta' \rightarrow \eta \pi^+ \pi^-)}
         {\Gamma(\eta' \rightarrow \eta \pi^0 \pi^0)} =  2.1 \pm 0.4 .
\end{equation}
Although isospin violating corrections seem to be small for this decay, we will not
modify our approach and keep different up- and down-quark masses.
The experimentally measured slope parameters are
\begin{equation} \label{eq:exp2}
\begin{array}{llll}
\mbox{Kalbfleisch \cite{Kalbfleisch:1974ku}:} \qquad &a= -0.16\pm 0.06,& & \\
\mbox{Alde et al. \cite{Alde:1986nw}:}  \qquad &a= -0.116\pm 0.026,\quad&b= a^2/4 
\pm 0.13^2,\quad&c=0.00\pm 0.03 \\
\mbox{Briere et al. \cite{Briere:1999bp}:}  \qquad &a= -0.042\pm 0.050 \footnote{The 
linear fit in Fig.~4 of 
\cite{Briere:1999bp}, however, shows a slope of $\alpha=-0.038$, in
contradiction with the value quoted and closer
to the value of \cite{Alde:1986nw}.},&& 
\end{array}
\end{equation}

\subsection{Tree diagram contributions to $\eta' \rightarrow \eta \pi \pi$}
\label{sec:etapeta2pitree}

The tree level amplitudes are given by
\begin{equation} 
A_{\eta' \eta + -}(s,t,u)=
A_{\eta' \eta 0 0}(s,t,u)=
\frac{4\sqrt{2}\,m^2_\pi \cvtwid{2}{1}}{3F^4}
\end{equation}
They lead to the decay rates and slope parameters
\begin{equation}
\begin{array}{llll} 
\Gamma(\eta' \rightarrow \eta \pi^0 \pi^0)  = 6.9 \keV ,&
a= -0.204,\quad& b= -0.005 ,\quad& c=-0.060;\\
\Gamma(\eta' \rightarrow \eta \pi^+ \pi^-)  = 12.2 \keV,\qquad&
a= -0.194 ,&  b= -0.005 , & c=-0.055.
\end{array}
\end{equation}
%
Clearly, we are not able to explain the data just by taking into account the tree level contribtutions
and need to include unitary corrections as for the other decays.

\subsection{Final state interactions in $\eta' \rightarrow \eta \pi \pi$}
\label{sec:etapeta2pifsi}

By iterating the effective $s$-wave potentials from meson meson scattering
we arrive at the decay rates
\begin{equation}
\begin{array}{llll} 
\Gamma(\eta' \rightarrow \eta \pi^0 \pi^0)  = 43.8 \keV ,&
a= -0.105,\quad& b= -0.065 ,\quad& c=-0.004;\\
\Gamma(\eta' \rightarrow \eta \pi^+ \pi^-)  = 77.7 \keV,\qquad&
a= -0.093 ,&  b= -0.059 , & c=-0.003
\end{array}
\end{equation}
being in qualitative agreement with experiment 
and with a relatively fixed ratio of $1.8$.

\subsection{Resonances  in $\eta' \rightarrow \eta \pi \pi$}

We can turn on final state interactions 
as described in Sec. \ref{sec:eta3pires} in both
the $s$ and $t,u$ channels independently 
to investigate the importance of resonances.
The results for the neutral decay read
\begin{equation}
\begin{array}{lllll}
\mbox{tree:}&
\Gamma= 6.9 \keV , &
a= -0.204 ,&  b= -0.005 ,& c=-0.060
\\
\mbox{tree}+s\mbox{-channel:}&
\Gamma  = 10.0 \keV,\quad & a= -0.455 ,\quad& b= -0.064 , \quad& c=-0.050\\
\mbox{tree}+t,u\mbox{-channel:}&\Gamma  =37.8 \keV , &
a= +0.021  , &  b= -0.005 , & c=+0.008
\\
\mbox{tree}+s,t,u\mbox{-channel:}\qquad&\Gamma  = 43.8 \keV , &
a= -0.105 , &  b= -0.065 , & c=-0.004
\end{array}
\end{equation}
and for the charged decay
\begin{equation}
\begin{array}{lllll}
\mbox{tree:}&
\Gamma= 12.2 \keV , &
a= -0.194 ,&  b=  -0.005 ,& c=-0.055
\\
\mbox{tree}+s\mbox{-channel:}&
\Gamma  = 17.9 \keV,\quad & a= -0.424 ,\quad& b=-0.059 , \quad& c=-0.043\\
\mbox{tree}+t,u\mbox{-channel:}&\Gamma  =66.2 \keV , &
a=+0.021  , &  b= -0.004, & c=+0.006
\\
\mbox{tree}+s,t,u\mbox{-channel:}\qquad&\Gamma  = 77.7 \keV , &
a= -0.093, &  b= -0.059 , & c=-0.003 .
\end{array}
\end{equation}
The main contribution to this decay arises from the final state interactions
in the $t,u$-channels, i.e. when the $\eta'$ decays into a pion and two mesons
which then rescatter infinitely many times to finally pass into a pion and an eta.
The iteration of the effective potentials in these channels mimics the effect of the
$a_0(980)$ which is usually generated within chiral unitary approaches, see \cite{Oller:1997ti}.
We therefore confirm the importance of the $a_0$ for this decay as it has been
claimed, e.g., in \cite{Fariborz:1999gr}. While in the tree level approach of
\cite{Fariborz:1999gr} the $a_0$ has been included as an explicit degree of freedom,
it automatically results in our approach by considering the final state interactions.
At first sight, these findings may seem to be contradictory to the conclusions
in \cite{Beisert:2002ad}, where the importance of the coupling constant
combination $\beta_0^{(0)}+ \beta_3^{(0)}$ was stressed
which dominates this decay via a contact interaction. However, in the framework
of \cite{Beisert:2002ad} effects of the $a_0$ 
are hidden in $\beta_0^{(0)}+ \beta_3^{(0)}$ via resonance saturation \cite{Ecker:1989te} 
and contribute indirectly through the contact interaction of $\beta_0^{(0)}+ \beta_3^{(0)}$.

\section{Conclusions}\label{sec:concl}

In the present work, we studied the hadronic decays
$\eta \rightarrow \pi \pi \pi$, $\eta' \rightarrow \pi \pi \pi$ and
$\eta' \rightarrow \eta \pi \pi$. To this end, the most general chiral effective
Lagrangian which describes the interactions between members of the lowest lying
pseudoscalar nonet $(\pi, K, \eta, \eta')$ was presented up to fourth chiral order.
The pertinent tree diagram amplitudes derived from this effective Lagrangian
are calculated. In all cases, the results for the decay rates are in sharp
disagreement with the measured values and suggest that unitary corrections are
dominant in these decays.

In order to improve the theoretical situation, we included final state interactions.
They were modelled by deriving the effective $s$-wave potentials for meson meson scattering
from the chiral effective Lagrangian and iterating them in a Bethe-Salpeter
equation.
The underlying idea is that the initial particle, $\eta$ or $\eta'$,
decays via chirally constrained vertices derived from the effective Lagrangian
into three mesons and that two out of these three mesons rescatter. There are three possible
ways in combining two of the mesons to a pair while leaving the third one unaffected
which corresponds to the $s$-, $t$- and $u$-channel, respectively.
The iteration of the effective potentials in these channels mimics the effect
of resonances and may be used to confirm calculations in which the same resonances
have been put in by hand.
With only a small set of parameters we are able to explain both rates
and spectral shapes of these decays.

The decay $\eta \rightarrow \pi \pi \pi$ violates isospin and omitting
the small contributions from electromagnetic interactions it can only
take place due to a finite mass difference $m_u-m_d$. 
The amplitude is then proportional to $m_u-m_d$ and provides
a sensitive probe on $SU(2)$ symmetry breaking.
Therefore, the $\pi^0 \eta \eta'$
mixing for different up- and down-quark masses $m_u \ne m_d$ is discussed in detail
as it arises from the second and fourth order Lagrangian, respectively.
Our results are in reasonable agreement with experiment after employing Dashen's theorem.
Conversely, this decay may be used to constrain the double quark mass ratio $Q$
and  we obtain $Q = 23.4 \pm 0.8$, 
a value which is in agreement with the requirement by Dashen's theorem and
indicates that higher order corrections to this low-energy theorem may be small.

For the decay $\eta' \rightarrow \pi \pi \pi$ which also violates isospin we obtain
agreement with the available data. Our result for the neutral mode,
$\Gamma(\eta' \rightarrow \pi^0 \pi^0 \pi^0)  = 291\eV$, is within the 
experimental error ranges
$\Gamma(\eta' \rightarrow \pi^0 \pi^0 \pi^0) = (311 \pm 77)\eV$,
while we can even deliver a prediction for the charged mode,
$\Gamma(\eta' \rightarrow \pi^0 \pi^+ \pi^-) = 294 \eV $, for which only an upper
experimental limit of $1005 \eV$ exists. The slope parameters for the spectral shapes
of the Dalitz plots may also serve as a check of our method.

The third and final decay we discussed is $\eta' \rightarrow \eta \pi \pi$.
Again, reasonable agreement with data was achieved and we could furthermore confirm within our approach
the importance of the $a_0(980)$ resonance for this decay as has been claimed in the
literature. This is also consistent with the findings in
\cite{Beisert:2002ad}, where effects of the $a_0$ 
are hidden  via resonance saturation 
in the contact interaction of the coupling constant combination
$\beta_0^{(0)}+ \beta_3^{(0)}$ which dominates the decay.

We were able to show that the results are in overall good agreement with the
decay rates, although the approach includes only a few parameters. In the case of the slope
parameters we are also in the right ballpark, but the experimental situation is less 
settled here and one needs new experiments with higher statistics in order to reduce
the error bars. Since we included only $s$-waves in our approach, we cannot expect, of course,
to reproduce the spectral shapes of the decays very precisely. The $p$-waves, although
small in magnitude, will alter our results for the Dalitz parameters,
and need to be included in order to make more accurate statements.

The method introduced in the present work has also set the stage for 
investigation on  further $\eta$ and $\eta'$ decays. Among these are
the anomalous decays $\eta/\eta' \rightarrow \gamma \gamma$ and 
$\eta/\eta' \rightarrow \pi^+ \pi^- \gamma$, the decays $\eta/\eta' \rightarrow \pi^0 \gamma \gamma$,
and the strong CP violating decays
$\eta/\eta' \rightarrow \pi \pi $.

\appendix

\section{Amplitudes}

In the appendix we present the bare 
$s$-wave amplitudes $A$
that are iterated in the Bethe-Salpeter equation
\eqref{eq:BSEMatrix} and 
related to the four-point functions at the tree level by equations analogous to
\eqref{eq:translate}.
We present here for brevity only the contributions from the leading order Lagrangian
$\Lagr^{(2)}$. Contributions from $\Lagr^{(4)}$ are treated in a similar way.

\subsection{Uncharged channels}

The amplitude matrix $A^0$ for the nine uncharged channels consists of an 
isospin conserving part
$A^{0}_{\epsilon^0}$ and an isospin breaking part $A^{0}_{\epsilon^1}$, i.e.
$A^0=A^{0}_{\epsilon^0}+A^{0}_{\epsilon^1}$.
The matrices $A^0_{\epsilon^0}$ and $A^0_{\epsilon^1}$ can be further decomposed as
\begin{eqnarray}
&A^{0}_{\epsilon^i}=\left(\begin{array}{c|c}
B^{0}_{\epsilon^i} & C^{0}_{\epsilon^i} \\[0.1cm] \hline 
(C^{0}_{\epsilon^i})^{T} & D^{0}_{\epsilon^i}
\end{array}\right) \qquad i=0,1 .
\end{eqnarray}
The submatrices $B$ and $D$ describe the couplings between the channels
($\pi^0\pi^0$, $\pi^+\pi^-$,$\eta\pi^0$, $\eta\eta$,$K^0\bar K^0$, $K^+K^-$)
and ($\eta'\pi^0$, $\eta'\eta$, $\eta'\eta'$), respectively. Transitions between
the first channels and the latter ones are given by the matrix $C$ and its transpose.
The matrices $B,C$ and $D$ read 
(with 
$\tilde v_2^{(3)}=\sfrac{1}{4}\decay^2-\sfrac{3}{2}\sqrt{6}v_3^{(1)}-9v_2^{(2)}+\sfrac{9}{2}\sqrt{6}v_3^{(3)}$ and 
$\tilde v_2^{(4)}=\sfrac{1}{4}\decay^2-2\sqrt{6}v_3^{(1)}+18\sqrt{6}v_3^{(3)}+54v_2^{(4)}$)
\begin{eqnarray}
&B^{0}_{\epsilon^0}=\left(\begin{array}{cccccc}
\frac{-\mmass{\pi}}{2F^2}&\frac{\mmass{\pi} - s}{\sqrt{2}F^2}&0&\frac{-\mmass{\pi}}{6F^2}&\frac{-s}{4\sqrt{2}F^2}&\frac{-s}{4\sqrt{2}F^2}\\
\frac{\mmass{\pi} - s}{{\sqrt{2}}F^2}&\frac{-s}{2F^2}&0&\frac{-\mmass{\pi}}{3{\sqrt{2}}F^2}&\frac{-s}{4F^2}&\frac{-s}{4F^2}\\
0&0&\frac{-\mmass{\pi}}{3F^2}&0&\frac{3s-4\mmass{K}}{4{\sqrt{3}}F^2}&\frac{4\mmass{K}-3s}{4{\sqrt{3}}F^2}\\
\frac{-\mmass{\pi}}{6F^2}&\frac{-\mmass{\pi}}{3{\sqrt{2}}F^2}&0&\frac{\mmass{\pi}-4\mmass{\eta}}{6F^2}&\frac{8\mmass{K}-9s}{12\sqrt{2}F^2}&\frac{8\mmass{K}-9s}{12\sqrt{2}F^2}\\
\frac{-s}{4\sqrt{2}F^2}&\frac{-s}{4F^2}&\frac{3s-4\mmass{K}}{4{\sqrt{3}}F^2}&\frac{8\mmass{K}-9s}{12{\sqrt{2}}F^2}&\frac{-s}{2F^2}&\frac{-s}{4F^2}\\
\frac{-s}{4\sqrt{2}F^2}&\frac{-s}{4F^2}&\frac{4\mmass{K}-3s}{4{\sqrt{3}}F^2}&\frac{8\mmass{K}-9s}{12{\sqrt{2}}F^2}&\frac{-s}{4F^2}&\frac{-s}{2F^2}
\end{array}\right)\\[0.5cm]\nonumber&
B^0_{\epsilon^1}={\displaystyle\frac{\masseps}{\mmass{\eta}-\mmass{\pi}}}\left(\begin{array}{cccccc}
0&0&\frac{\mmass{\eta}-\mmass{\pi}}{\sqrt{6}F^2}&0&\frac{3s-4\mmass{K}}{6\sqrt{2}F^2}&\frac{4\mmass{K}-3s}{6\sqrt{2}F^2}\\
0&0&\frac{3s-4\mmass{\pi}}{3{\sqrt{3}}F^2}&0&0&0\\
\frac{\mmass{\eta}-\mmass{\pi}}{{\sqrt{6}}F^2}&\frac{3s-4\mmass{\pi}}{3{\sqrt{3}}F^2}&0&\frac{\mmass{\pi}-\mmass{\eta}}{\sqrt{6}F^2}&\frac{4\mmass{\pi}-3s}{6{\sqrt{3}}F^2}&\frac{4\mmass{\pi}-3s}{6{\sqrt{3}}F^2}\\
0&0&\frac{\mmass{\pi}-\mmass{\eta}}{{\sqrt{6}}F^2}&0&\frac{5\mmass{\eta}-\mmass{\pi}-3s}{6\sqrt{2}F^2}&\frac{3s-5\mmass{\eta}+\mmass{\pi}}{6\sqrt{2}F^2}\\
\frac{3s-4\mmass{K}}{6\sqrt{2}F^2}&0&\frac{4\mmass{\pi}-3s}{6{\sqrt{3}}F^2}&\frac{5\mmass{\eta}-\mmass{\pi}-3s}{6{\sqrt{2}}F^2}&0&0\\
\frac{4\mmass{K}-3s}{6\sqrt{2}F^2}&0&\frac{4\mmass{\pi}-3s}{6\sqrt{3}F^2}&\frac{3s-5\mmass{\eta}+\mmass{\pi}}{6\sqrt{2}F^2}&0&0
\end{array}\right)
\end{eqnarray}
\bigskip

\begin{eqnarray}
&C^{0}_{\epsilon^0}=\left(\begin{array}{ccc}
0&\frac{-4\mmass{\pi}\cvtwid{2}{1}}{3F^4}&\frac{6(s-2\mmass{\pi})\coeffv{1}{2}-4\mmass{\pi}\cvtwid{2}{2}}{3F^4}\\ 
0&\frac{-8\mmass{\pi}\cvtwid{2}{1}}{3\sqrt{2}F^4}&\frac{12(s-2\mmass{\pi})\coeffv{1}{2}-8\mmass{\pi}\cvtwid{2}{2}}{3\sqrt{2}F^4}\\
\frac{-8\mmass{\pi}\cvtwid{2}{1}}{3\sqrt{2}F^4}&0&0\\
0&\frac{(8\mmass{\eta}-4\mmass{\pi})\cvtwid{2}{1}}{3F^4}&\frac{6(s-2\mmass{\eta})\coeffv{1}{2}-4\mmass{\eta}\cvtwid{2}{2}}{3F^4}\\
\frac{2(\mmass{\eta}+\mmass{\pi})\cvtwid{2}{1}}{\sqrt{6}F^4}&\frac{2(3\mmass{\eta}-\mmass{\pi})\cvtwid{2}{1}}{3\sqrt{2}F^4}&\frac{12(s-2\mmass{K})\coeffv{1}{2}-8\mmass{K}\cvtwid{2}{2}}{3\sqrt{2}F^4}\\
\frac{-2(\mmass{\eta}+\mmass{\pi})\cvtwid{2}{1}}{\sqrt{6}F^4}&\frac{2(3\mmass{\eta}-\mmass{\pi})\cvtwid{2}{1}}{3{\sqrt{2}}F^4}&\frac{12(s-2\mmass{K})\coeffv{1}{2}-8\mmass{K}\cvtwid{2}{2}}{3{\sqrt{2}}F^4}
\end{array}\right)
\nonumber\\[0.5cm]&%
C^{0}_{\epsilon^1}={\displaystyle\frac{\masseps}{\mmass{\eta}-\mmass{\pi}}}\left(\begin{array}{ccc}
\frac{4(\mmass{\eta}-2\mmass{\pi})\cvtwid{2}{1}}{{\sqrt{3}}F^4}&0&0\\ 
\frac{8(\mmass{\eta}-2\mmass{\pi})\cvtwid{2}{1}}{3\sqrt{6}F^4}&0&0\\
0&\frac{8\mmass{\eta}\cvtwid{2}{1}}{\sqrt{6}F^4}&0\\
\frac{4\mmass{\eta}\cvtwid{2}{1}}{\sqrt{3}F^4}&0&0\\
\frac{2(7\mmass{\eta}-5\mmass{\pi})\cvtwid{2}{1}}{3\sqrt{6}F^4}&\frac{-2(\mmass{\eta}+\mmass{\pi})\cvtwid{2}{1}}{3\sqrt{2}F^4}&\frac{-4(\mmass{\eta}-\mmass{\pi})(3\coeffv{1}{2}+\cvtwid{2}{2})}{3\sqrt{2}F^4}\\
\frac{2(7\mmass{\eta}-5\mmass{\pi})\cvtwid{2}{1}}{3\sqrt{6}F^4}&\frac{2(\mmass{\eta}+\mmass{\pi})\cvtwid{2}{1}}{3\sqrt{2}F^4}&\frac{4(\mmass{\eta}-\mmass{\pi})(3\coeffv{1}{2}+\cvtwid{2}{2})}{3{\sqrt{2}}F^4}
\end{array}\right)
\end{eqnarray}
\bigskip

\begin{eqnarray}
&D^{0}_{\epsilon^0}=\left(\begin{array}{ccc}
\frac{-6\coeffv{1}{2}(\mmass{\eta'}-\mmass{\pi}-s)^2-8\mmass{\pi}s\cvtwid{2}{2}}{3F^4s}\hspace{-1cm}&0&0\\
0&\frac{-6\coeffv{1}{2}(\mmass{\eta'}-\mmass{\eta}-s)^2-8\mmass{\eta}s\cvtwid{2}{2}}{3F^4s}&\frac{4(\mmass{\eta}-\mmass{\pi})\cvtwid{2}{3}}{3F^4}\\
0&\frac{4(\mmass{\eta}-\mmass{\pi})\cvtwid{2}{3}}{3F^4}&\hspace{-0.5cm}\frac{36\coeffv{0}{4}-24\mmass{\eta'}\coeffv{1}{2}+72\mmass{\eta'}\coeffv{4}{2}-2\cvtwid{2}{4}(\mmass{\eta}+\mmass{\pi})}{3F^4}
\end{array}\right)
\nonumber\\[0.5cm]&%
D^{0}_{\epsilon^1}={\displaystyle\frac{\masseps}{\mmass{\eta}-\mmass{\pi}}}\left(\begin{array}{ccc}
0&0&\frac{4(\mmass{\eta}-\mmass{\pi})\cvtwid{2}{3}}{\sqrt{3}F^4}\\
0&0&0\\
\frac{4(\mmass{\eta}-\mmass{\pi})\cvtwid{2}{3}}{{\sqrt{3}}F^4}&0&0
\end{array}\right)
\end{eqnarray}

\subsection{Charged channels}
For the four charged channels ($\pi^0\pi^+$, $\eta\pi^+$, $K^+\bar K^0$, $\eta'\pi^+$),
the full matrix $A^+$ is given by 
\begin{equation}
A^+=\left(\begin{array}{cccc}
\frac{s-2\mmass{\pi}}{2F^2}&
\frac{\masseps(4\mmass{K}-3s)}{6{\sqrt{3}}F^2(\mmass{\eta}-\mmass{\pi})}&
\frac{\masseps(4\mmass{K}-3s)}{6{\sqrt{2}}F^2(\mmass{\eta}-\mmass{\pi})}&
\frac{8\masseps (\mmass{\eta}-2\mmass{\pi})\cvtwid{2}{1}}{3\sqrt{6}F^4(\mmass{\eta}-\mmass{\pi})}
\\
\frac{\masseps ( 4\mmass{K} - 3s ) }{6{\sqrt{3}}F^2( \mmass{\eta} - \mmass{\pi} ) }&
\frac{-\mmass{\pi}}{3F^2}&
\frac{4\mmass{K}-3s}{2{\sqrt{6}}F^2}&
\frac{-4{\sqrt{2}}\mmass{\pi}\cvtwid{2}{1}}{3F^4}
\\
\frac{\masseps(4\mmass{K}-3s)}{6\sqrt{2}F^2(\mmass{\eta}-\mmass{\pi})}&
\frac{4\mmass{K}-3s}{2\sqrt{6}F^2}&
\frac{-s}{4F^2}&
\frac{-2(\mmass{\eta}+\mmass{\pi})\cvtwid{2}{1}}{\sqrt{3}F^4}
\\
\frac{8\masseps ( \mmass{\eta} - 2\mmass{\pi} ) \cvtwid{2}{1}}{3\sqrt{6}F^4 ( \mmass{\eta} - \mmass{\pi} ) }&
\frac{-4{\sqrt{2}}\mmass{\pi}\cvtwid{2}{1}}{3F^4}&
\frac{-2( \mmass{\eta} + \mmass{\pi} ) \cvtwid{2}{1}}{{\sqrt{3}}F^4}&
\frac{-2\coeffv{1}{2}(\mmass{\eta'}-\mmass{\pi}-s)^2}{F^4s}-\frac{8\mmass{\pi}\cvtwid{2}{2}}{3F^4}
\end{array}\right)
\end{equation}


\bibliography{eta3pi} 
\bibliographystyle{eta3pi}

\end{document}